\documentclass[prb,aps,twocolumn,groupedaddress,floats,final,nopacks]{revtex4-2}
\usepackage{graphicx}
\usepackage{subfigure}
\usepackage{amssymb}
\usepackage{latexsym}
\usepackage{epsfig}
\usepackage{psfrag}
\usepackage{dcolumn}
\usepackage{amsmath,amsfonts,amssymb,bm,ulem}
\usepackage{color}
\usepackage{float}

\newcommand{\be}{\begin{equation}}
\newcommand{\ee}{\end{equation}}
\newcommand{\bea}{\begin{eqnarray}}
\newcommand{\eea}{\end{eqnarray}}

\begin{document}
\title{Solving the Bethe-Salpeter Equation in Real Frequencies at Finite Temperature}

\author{I.S. Tupitsyn}
\affiliation{Department of Physics, University of Massachusetts, Amherst, MA 01003, USA}
\author{N.V. Prokof'ev}
\affiliation{Department of Physics, University of Massachusetts, Amherst, MA 01003, USA}

%\date{\today}

\begin{abstract}
Self-consistent Hartree-Fock approximation combined with solutions of the Bethe-Salpeter equation offers a powerful tool for studies of strong correlation effects arising in condensed matter models, nuclear physics, quantum field theories, and real materials. The standard finite-temperature approach would be to first solve the problem in the Matsubara representation and then apply numerical analytic continuation to the real-frequency axis to link theoretical results with experimental probes, but this ill-conditioned procedure often distorts important spectral features even for very accurate imaginary-frequency data. We demonstrate that the ladder-type finite-temperature Bethe-Salpeter equation in the Hartree-Fock basis for the 3-point vertex function and, ultimately, system's polarization can be accurately solved directly on the real frequency axis using the diagrammatic Monte Carlo technique and series resummation. We illustrate the method by applying it to the homogeneous electron gas model and demonstrate how multiple scattering events renormalize Landau damping.
\end{abstract}

\maketitle

%%%%%%%%%%%%%%%%%%%%%%%%%%%%%%%%%%%%%%%%%%%%%%%%%%%%%%%%%%%%%%%%%%%%%%%%%%%%%%%%%%

\section{Introduction} 

The Bethe-Salpeter equation (BSE) was initially introduced to study two-particle Green's function and the corresponding bound states in nuclear physics \cite{BS1951}. Later on it was formulated for studies of optical spectra in solids \cite{Hanke1979,Benedict1998} and molecules \cite{Louie1989,Rohlfing2012}. In material science/quantum chemistry methods one often starts by establishing the Hartree-Fock (HF) basis and proceeds with the electronic structure calculations and solutions of the BSE for two-particle Green's function and related properties \cite{Olevano2017,leng2016,vissher2022}. In strongly correlated models and quantum field theories BSE is used to calculate the effect of  vertex corrections \cite{mahan3rd,hedin1965}. This is especially important in the case of polarization when looking at charge and spin responses because neglecting vertex corrections leads to wrong results \cite{TTKP2021}.

The standard finite-temperature approach is to first solve the problem in the Matsubara domain \cite{Matsubara1955} and then apply a numeric analytic continuation procedure to obtain real frequency results for comparison with experimental observations. However, even if original data is known with high accuracy, the final step distorts, or even fails to resolve, important spectral features unless one imposes constrains on how the result is supposed  to behave \cite{Goulko2017}. Until recently, this problem standing in the way of accurate theoretical descriptions of experimentally relevant observables was considered unavoidable.

The breakthrough development done in the context of diagrammatic Monte Carlo (diagMC) for Fermi-Hubbard model in Refs.~\cite{LeBlanc2019,LeBlanc2020a,LeBlanc2020b} (see also Refs.~\cite{FerreroRT,Ferrero2020}) was that one can automatically express real-frequency answers for an arbitrary diagram using the so-called algorithmic Matsubara integration (AMI) protocol and, thus, completely eliminate the need for numeric analytic continuation. It works best for expansions in terms of ``bare'' (non-interacting) Green's functions and frequency-independent interactions. However, for strongly correlated regimes, when diagrammatic expansions in terms of original potentials diverge---a typical example would be the fundamental homogeneous electron gas (HEG), or jellium, model---expansions should be performed in terms of ``dressed" propagators and screened potentials, which include some interactions effects non-perturbatively. The corresponding generalization of the AMI protocol was described in Ref.~\cite{TTKP2021} but its implementation for high-order diagrammatic simulations poses significant technical challenges.

Current applications of the AMI protocol for jellium are based on expansion in terms of the Yukawa potential,
$Y=4 \pi e^2/(q^2 + \kappa^2)$, with optimally chosen screening momentum $\kappa$ and self-consistent HF single particle propagators, $G$, see Refs.~\cite{LCHPT2022,Chen2019,Haule2022} and the text below Eq.~(\ref{jellium}). In brief, the AMI technique lists all diagrams of order $N$ for an observable of interest, eliminates all sums over internal Matsubara frequencies for every listed diagram by an exact analytic transformation, and stochastically samples the remaining momentum integrals. However, it remains biased because all poles are regularized by adding small but finite imaginary part to the external frequency, $\Omega \to \Omega + i\eta$. The exact answer is recovered by taking the $\eta \to +0$ limit which requires extremely long computation times for high-order diagrams \cite{LCHPT2022}.

In this work, we apply the AMI and series resummation techniques for solving the ladder-type
finite-temperature BSE equation for the $3$-point vertex function $\Gamma^{(3)}$ convoluted with two Green's functions to obtain system's polarization directly on the real-frequency axis. We take the crucial new step of implementing the limit of $\eta \to +0$ explicitly; the same procedure may be used for any model with instantaneous interaction. The resulting scheme is not only numerically exact, it is also very efficient and allows one to reach expansion orders high-enough for performing reliable resummations of divergent series. The same expansion orders cannot be reached if simulations are performed with appropriately small finite-$\eta$ regularization.

For illustration, we compute the polarization of the HEG in the HF basis, and compare it to the random-phase-approximation (RPA) predictions (also in the HF basis) to see the effects of multiple particle-hole scattering processes. In particular, we find that the Landau damping coefficient is significantly increased relative to the RPA prediction.

\section{Model and simulation setup}

Hamiltonian of the HEG is defined by
\begin{equation}
H=\sum_i \frac{k_i^2}{2m} + \sum_{i<j} \frac{e^2}{|\mathbf{r}_i -\mathbf{r}_j|} - \mu N ,
\label{jellium}
\end{equation}
with $m$ the electron mass, $\mu$ the chemical potential, and $e$ the electron charge. We use the inverse Fermi momentum, $1/k_F$, and Fermi energy, $\varepsilon_F = k_F^2/2m$, as units of length and energy, respectively; the definition of the Coulomb parameter $r_s$ in terms of the particle number density, $\rho = k_F^3/3\pi^2$, and Bohr radius, $a_B=1/me^2$, is standard: $4\pi r_s^3/3 = 1/ \rho a_B^3$.
\begin{figure}[h]
%\vspace{-2mm}
\centerline{\includegraphics[angle = 0,width=0.98\columnwidth]{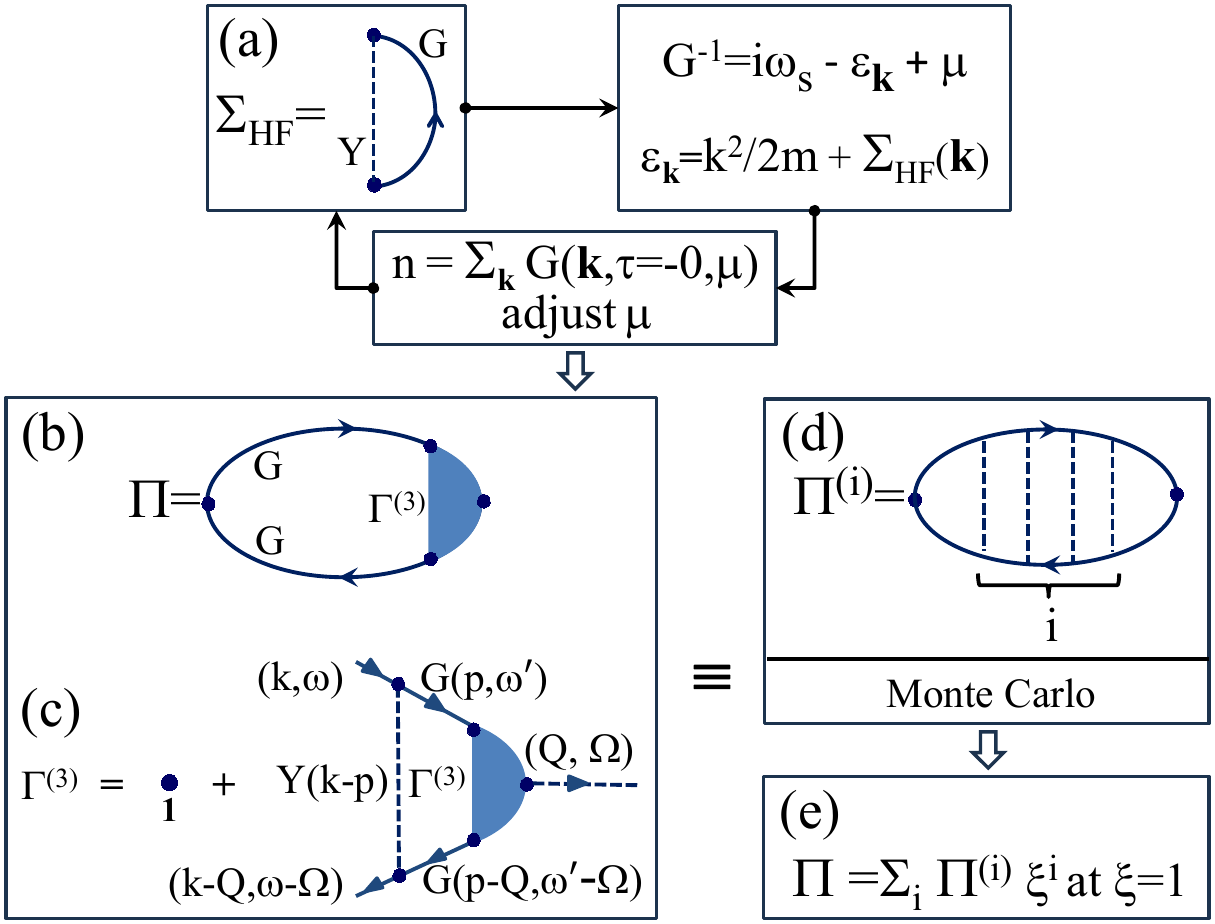}}
%\vspace{-2mm}
\caption{Flowchart of the technique. Calculation starts from the self-consistent HF
solution for the proper self-energy, $\Sigma_{HF}$ (a), with the chemical potential, $\mu$, constantly adjusted to keep the number density, $n$, fixed. Converged solution, $G$, is subsequently used for computing the system's polarization, $\Pi$ (b), from the series of ladder diagrams on the real frequency axis. The series can be written in terms of the BSE (c) for the dressed vertex function, $\Gamma^{(3)}$, or as Taylor series expansion in the number of interaction lines. Monte Carlo simulations in the $\eta \to 0$ limit are used to compute partial $i$-th order contributions (d). In all diagrams solid and dashed lines correspond to $G$ and Yukawa potential, respectively. }
\label{Fig1}
\end{figure}

For a meaningful diagrammatic expansion one has to work with screened interactions and counter-terms.
In this work we follow the setup formulated in Ref.~\cite{Chen2019} and expand in the Yukawa potential
on top of the self-consistent HF solution for the Green's function: $G^{-1} = G^{-1}_0 - \Sigma_{HF}[G]$,
where $G_0$ is the bare Green's function and $\Sigma_{HF}$ is the Fock proper self-energy shown in Fig.~\ref{Fig1}(a) (Hartree diagram is absent by charge neutrality). In the converged HF solution,
$G = (i\omega_s - k^2/2m - \Sigma_{HF}(\mathbf{k}) + \mu)^{-1} \equiv (i\omega_s - \epsilon_{k} +\mu)^{-1}$, the chemical potential is also self-consistently adjusted to keep the electron density fixed, see upper part of Fig.~\ref{Fig1}. By incorporating Fock diagrams into $G$ one simplifies the expansion by omitting all diagrams with Fock-type self-energy insertions.

Formally, not only the screening momentum $\kappa$ in the Yukawa potential, but the entire potential itself
is arbitrary as long as it does not feature divergent behavior at $q=0$. Any difference between the Coulomb,
$V(q)=4\pi e^2/q^2$, and expansion potentials is taken care of by a counter-term $ C(q)=1/V(q) - 1/Y(q)$ which keeps the field theoretical representation of the model exact \cite{ShiftAct,homotopy,Chen2019}. For example,
the physical screened interaction is given by $1/W = 1/Y - [\Pi - C(q)] \equiv 1/V - \Pi$ where $\Pi$ is the
system polarization. Ultimately, final results are supposed to show little dependence on the choice of $Y(q)$.
Flexibility in reformulating the diagrammatic expansion can, and should be, used for optimization of series
convergence properties. This is precisely how recent work reported in Refs.~\cite{Chen2019,Haule2022}
achieved unprecedented accuracy for single-particle and static linear response properties of HEG.

\begin{figure}[H]
\centerline{\includegraphics[angle = 0,width=0.48\columnwidth]{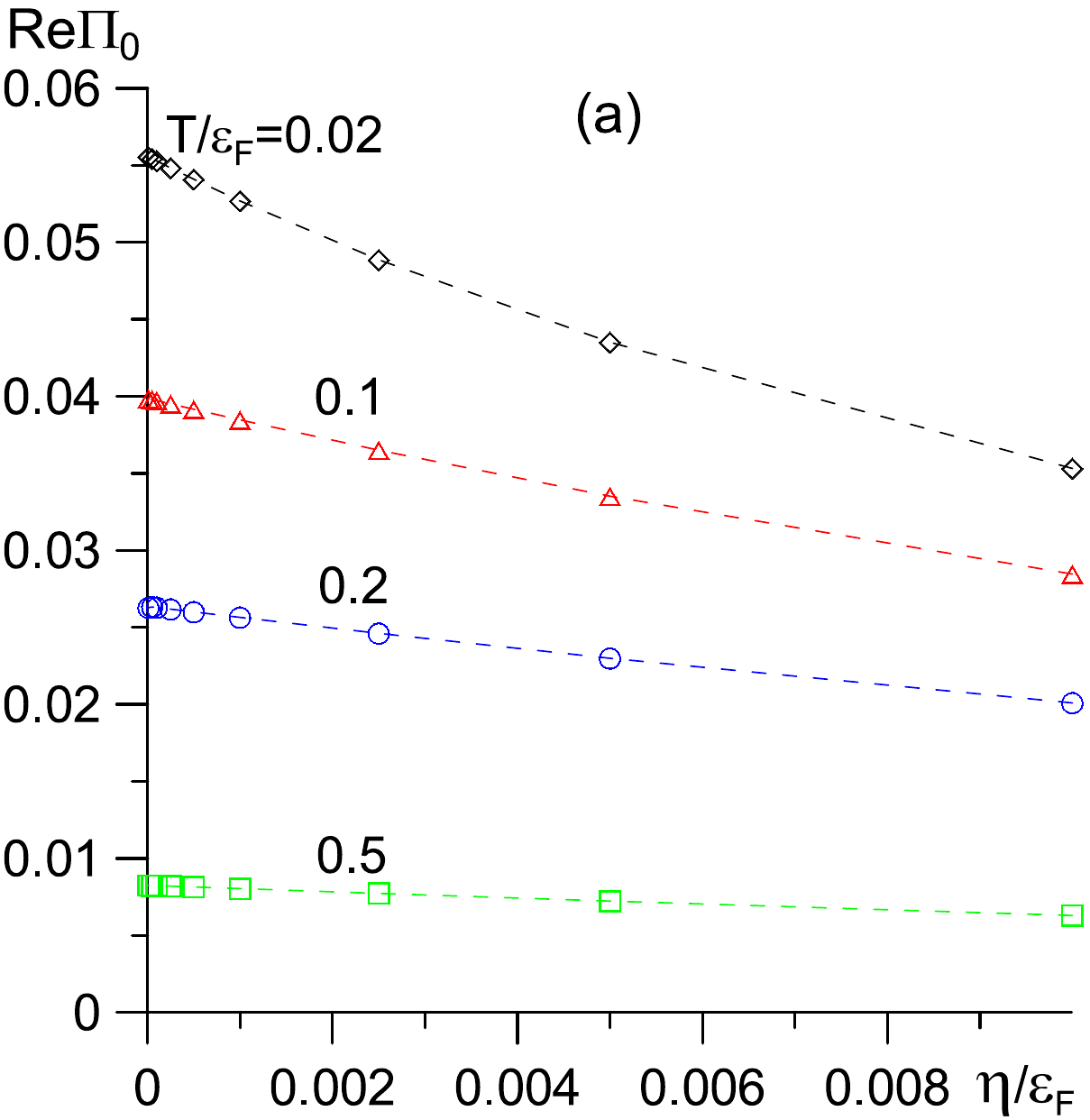}
            \includegraphics[angle = 0,width=0.48\columnwidth]{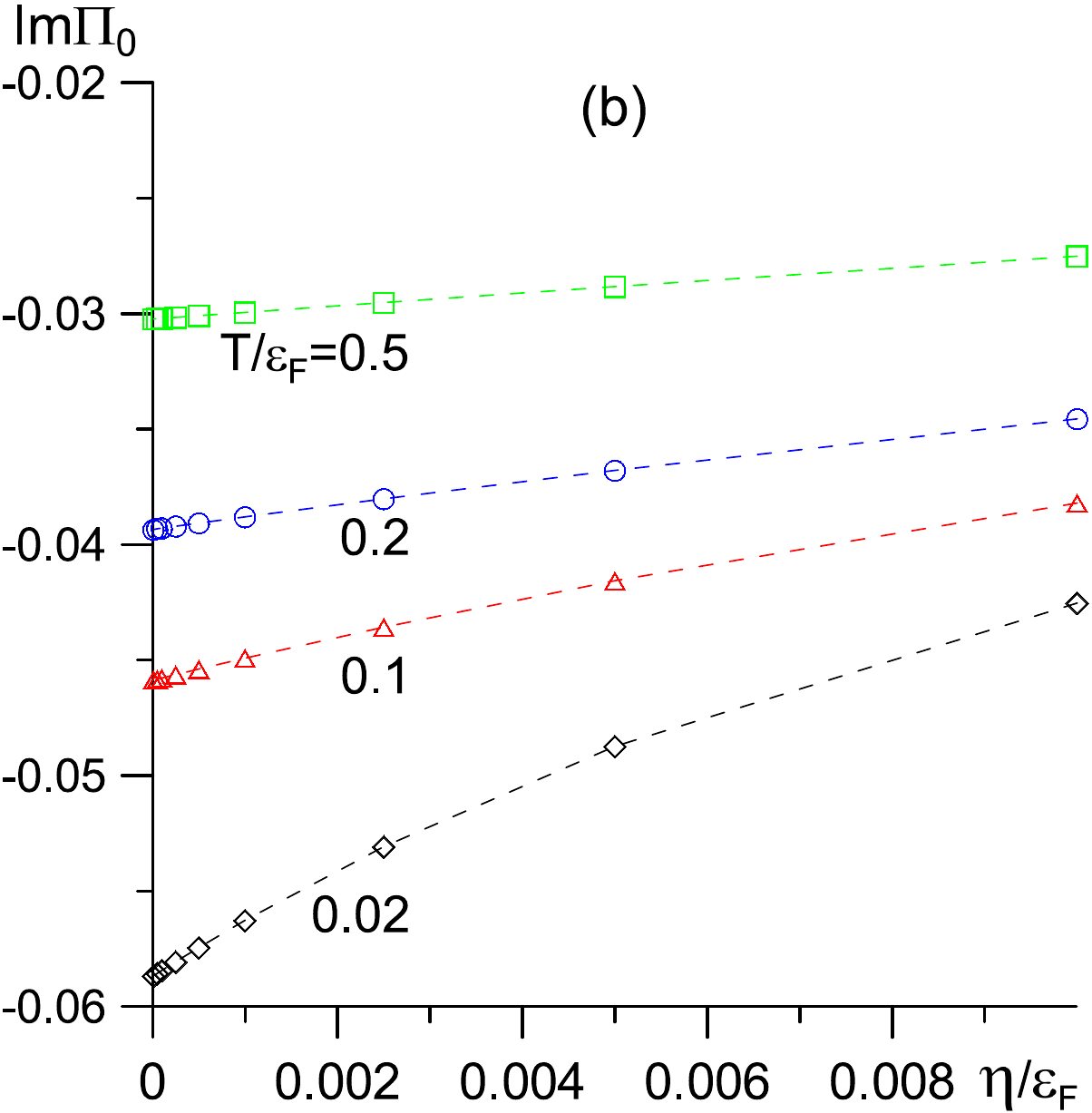}}
\caption{Real and imaginary parts of the polarization $\Pi_0$ in the HF basis (see Fig.\ref{Fig1}(b) with $\Gamma^{(3)}=1$) at $r_s=2$ for $Q/k_F=0.09844$ and $\Omega/\varepsilon_F=0.2$ as a function of $\eta$ (the smallest value of $\eta /\varepsilon_F$ here is $10^{-5}$). }
\label{Fig2}
\end{figure}

In this work we apply the above field-theoretical setup to explore how the polarization is modified
by multiple scattering of the particle-hole pair at finite temperature. Since the calculation is performed directly on the real frequency axis, our results can be immediately used to obtain the Landau damping parameter, see Ref.~\cite{LD_TP_2023}, or describe the so-call charge loss function, $\mathrm{Im} W$. The set of diagrams involved is presented in Figs.~\ref{Fig1} (b), (c), and (d) where the ladder-diagram series for polarization are first expressed in terms of the BSE for the dressed three-point vertex function $\Gamma^{(3)}$, see \ref{Fig1} (c), which is then convoluted with two Green's functions, see \ref{Fig1} (b). As noticed in Ref.~\cite{TTKP2021}, if the calculation is restricted to have no more than one ladder rung, the result for $\Pi = \Pi_0+\Pi_1$, where $\Pi_i$ is the contribution of the diagram with $i$ rungs Fig.~\ref{Fig1}(d), may violate causality. Also, an observation that $\Pi_1$ is not small compared to $\Pi_0$ raises the question of importance of contributions from high-order ladder diagrams. Summation of the entire ladder series is supposed to solve both problems.

Before proceeding, let us elaborate on the importance of considering very small values of $\eta $ when using it for regularization of poles. Even the lowest order diagram demonstrates significant finite-$\eta$ correction at low temperature, see Fig.~\ref{Fig2}. At $T/\varepsilon_F=0.02$ the finite-$\eta$ effect in $\Pi_0$ is as large as $20\%$ for $\eta /\varepsilon_F=0.005$. One may attempt to extrapolate results towards $\eta=0$, but this procedure is not guaranteed to work for all higher-order contributions because the $\eta$-dependence is not necessarily monotonic.

\noindent \textit{BSE in the $\eta \to +0$ limit.} As discussed above, physical results may not depend on the choice of $Y(q)$ for setting up the expansion. This, in particular, implies that good approximations should produce results nearly independent of the choice of the screening momentum in $Y(q)$; any significant dependence would be a clear sign of the approximation bias. This is precisely why $\Pi = \Pi_0+\Pi_1$ needs to be supplemented with additional diagrams which are supposed to restore causality and change the function shape. In what follows we demonstrate that summation of the entire ladder series produces results with relatively weak dependence on $\kappa^2$ when the latter is changed by more than a factor of two in the vicinity of the Thomas-Fermi momentum $\kappa_{TF}=6 \pi \rho e^2/ \varepsilon_F$. In this work we perform all calculations for $r_s=2$ when $\kappa_{TF} \approx 1.15 k_F$.

The $i$-th order contribution to the ladder diagram series for $\Pi$, see Figs.~\ref{Fig1}(b) and \ref{Fig1}(c), is given by (in what follows we treat $\Omega$ and ${\bf Q}$ as external parameters and do not mention them explicitly as function variables)
\begin{equation}
\Pi_i= - 2 \left( \prod_{j=1}^{i+1}\int \frac{d {\bf p}_{j}}{(2\pi)^3} \right)
\prod_{j=1}^{i+1} {\cal F}_{{\bf p}_j}
\prod_{j=1}^{i} Y({\bf p}_{j+1}-{\bf p}_j) \,,
\label{Pin}
\end{equation}
where
\begin{equation}
{\cal F}_{{\bf p}} =
\frac{n_{{\bf p} +{\bf Q}}-n_{\mathbf{p}}} {\Omega - \epsilon_{{\bf p} +{\bf Q}} + \epsilon_{\mathbf{p}} + i \eta},
\label{defF}
\end{equation}
with $n_{\mathbf{p}}=(e^{\epsilon_p -\mu}+1)^{-1}$. While this setup is immediately suitable for Monte Carlo
simulations with small finite $\eta$, it quickly becomes too expensive as $i$ increases. The reason is the severe
sign problem coming from Monte Carlo sampling of the principle part integrals around poles of $F$ functions.

This problem can be eliminated by taking the limit of $\eta \to +0$ explicitly with the help of the $\lim_{\eta \to +0} \frac{1}{x+i\eta} =-i\pi \delta (x) + {\cal P}\frac{1}{x}$ formula were ${\cal P}$ stands for the principal part value. In practice it means that integrals featuring simple poles can be transformed identically as
\begin{equation}
\int_{a}^{b} dx \frac{h(x)}{\Omega - g (x) +i\eta } =  \int_{a}^{b} dx
\left[ \alpha(x,x_0)+ i\gamma (x,x_0) \right]  \,,
\label{poles}
\end{equation}
where $x_0$ is the solution of the $g(x_0)=\Omega$ equation and
\begin{eqnarray}
\gamma (x,x_0)&=& -\pi \frac{h(x_0)}{|g'(x_0)|} N(x,x_0)
\;\; \mbox{if} \; x_0 \in(a,b), \nonumber \\
\gamma (x,x_0)&=&0 \qquad \qquad \qquad \qquad \;\;\; \mbox{otherwise},
\label{gamma}
\end{eqnarray}
\begin{eqnarray}
\alpha (x,x_0) &=& \frac{1}{2}\left[ \frac{h(x)}{\Omega - g (x)}+\frac{h(x_r)}{\Omega - g (x_r)} \right]
\;\; \mbox{if} \; (x_0,x_r) \in(a,b)  \nonumber \\
\alpha (x,x_0) &=& \frac{h(x)}{\Omega - g (x)} \qquad \qquad \qquad \qquad
\;\; \mbox{otherwise},
\label{alpha}
\end{eqnarray}
with $x_r=2x_0-x$.
Here $N(x,x_0)$ is an arbitrary function normalized to unity on the $(a,b)$ interval--it may explicitly depend on $x_0$. By construction, the $\alpha +i\gamma$ function is not singular at the pole location.
In our case, we choose variables $x = \cos \widehat{{\bf p} {\bf Q }}$ to regularize poles.
For a given spherically symmetric dispersion relation $\epsilon_{\mathbf{p}}$ the location of the pole $x_0 (p,Q,\Omega)$ as a function of $p$ is tabulated on a fine grid at the start of the simulation and subsequently is treated as a known function.

However, taking the limit $\eta \to +0$ also involves ``splitting" the contribution of the diagram with variable $x$ into up to three contributions from variables $x$, $x_0$, and $x_r$ if $(x_0, x_r) \in (a,b)$. As the ladder diagram order increases, one now faces the problem of computing up to $3^{i+1}$ contributions for some multi-dimensional momentum points $\{ {\bf p}_j \}$ because momentum variables are linked by potentials $Y$, see (\ref{Pin}). This remaining problem is solved by ``decoupling" momentum integrals using Fourier transforms of Yukawa potentials
\begin{equation}
Y({\bf p}_{j+1}-{\bf p}_j)= \int d{\bf r}_j Y(r_j) \, e^{i({\bf p}_{j+1}-{\bf p}_j)\cdot {\bf r}_j } \,,
\label{Fourier}
\end{equation}
where $Y(r)=(e^2/r)\,e^{-\kappa r}$. Now
\begin{equation}
\Pi_i = -2 \left( \prod_{j=1}^{i}\int d {\bf r}_{j} Y(r_j) \right)
\prod_{j=1}^{i+1}\int \frac{d {\bf p}_{j}}{(2\pi)^3}
\tilde{\cal F}_{{\bf p}_j, {\bf r}_{j-1}-{\bf r}_{j}}
\label{PinF}
\end{equation}
with
\begin{equation}
\tilde{\cal F}_{{\bf p}, {\bf R}} = e^{i{\bf R} \cdot {\bf p} }
\frac{n_{{\bf p} +{\bf Q}}-n_{\mathbf{p}}} {\Omega - \epsilon_{{\bf p} +{\bf Q}} + \epsilon_{\mathbf{p}} + i \eta}.
\label{defFt}
\end{equation}
Here we assume that ${\bf r}_0 = {\bf r}_{i+1} =0$ for convenience of compact notations. According to (\ref{PinF}), for a given set of integration variables the integrand is a product of functions which depend only on one momentum variable. When the $\eta \to +0$ limit is taken, each $\tilde{\cal F}$ function may be split into up to three contributions, but the computational cost of evaluating the total contribution is now reduced to $3(i+1)$ instead of $3^{i+1}$. Additional real space integrals do not pose any technical or efficiency problems for Monte Carlo simulations, which all can be performed on a single laptop.

\section{Series Convergence} 

\begin{figure}[t]
\centerline{\includegraphics[angle = 0,width=0.48\columnwidth]{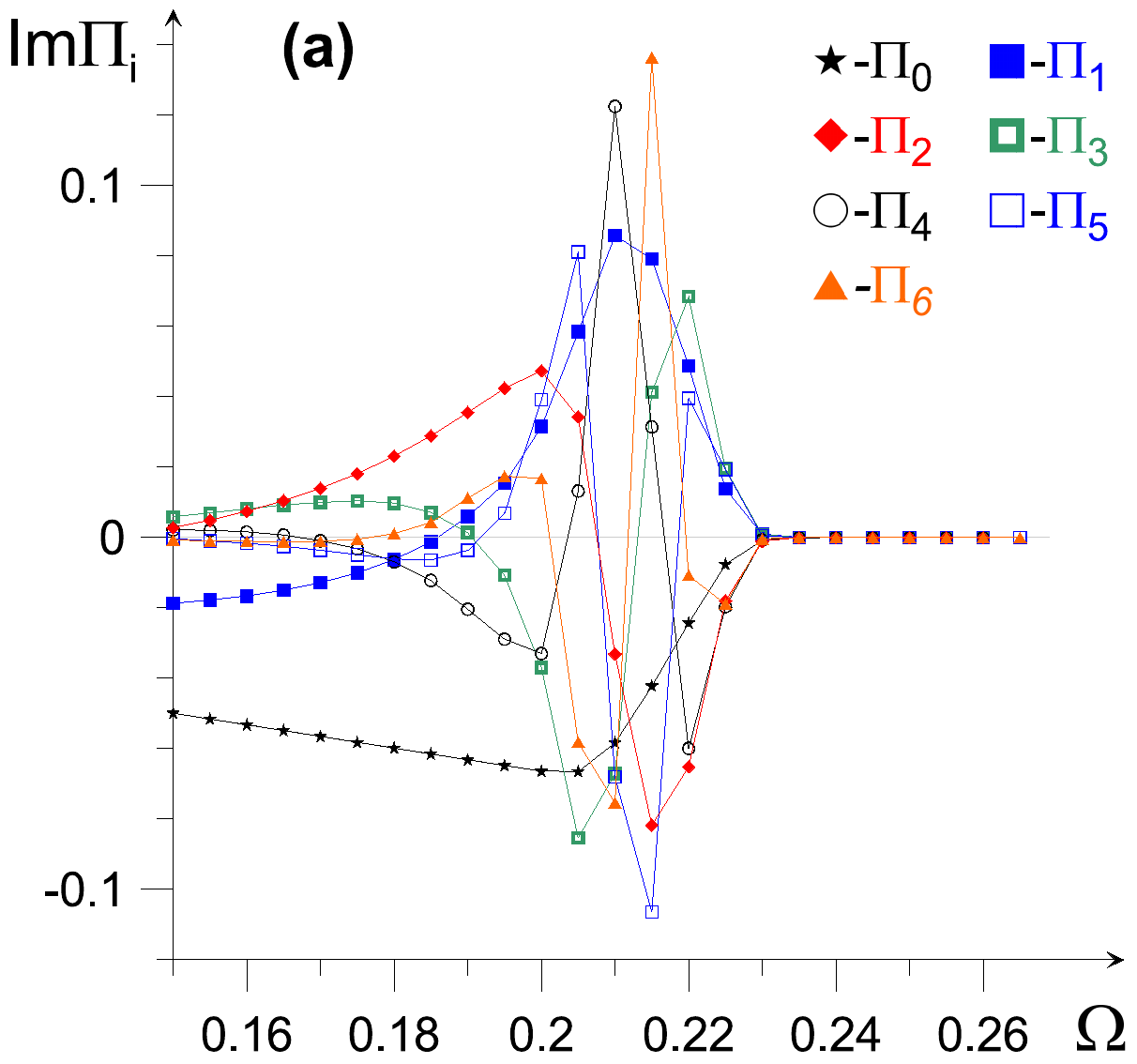}
            \includegraphics[angle = 0,width=0.48\columnwidth]{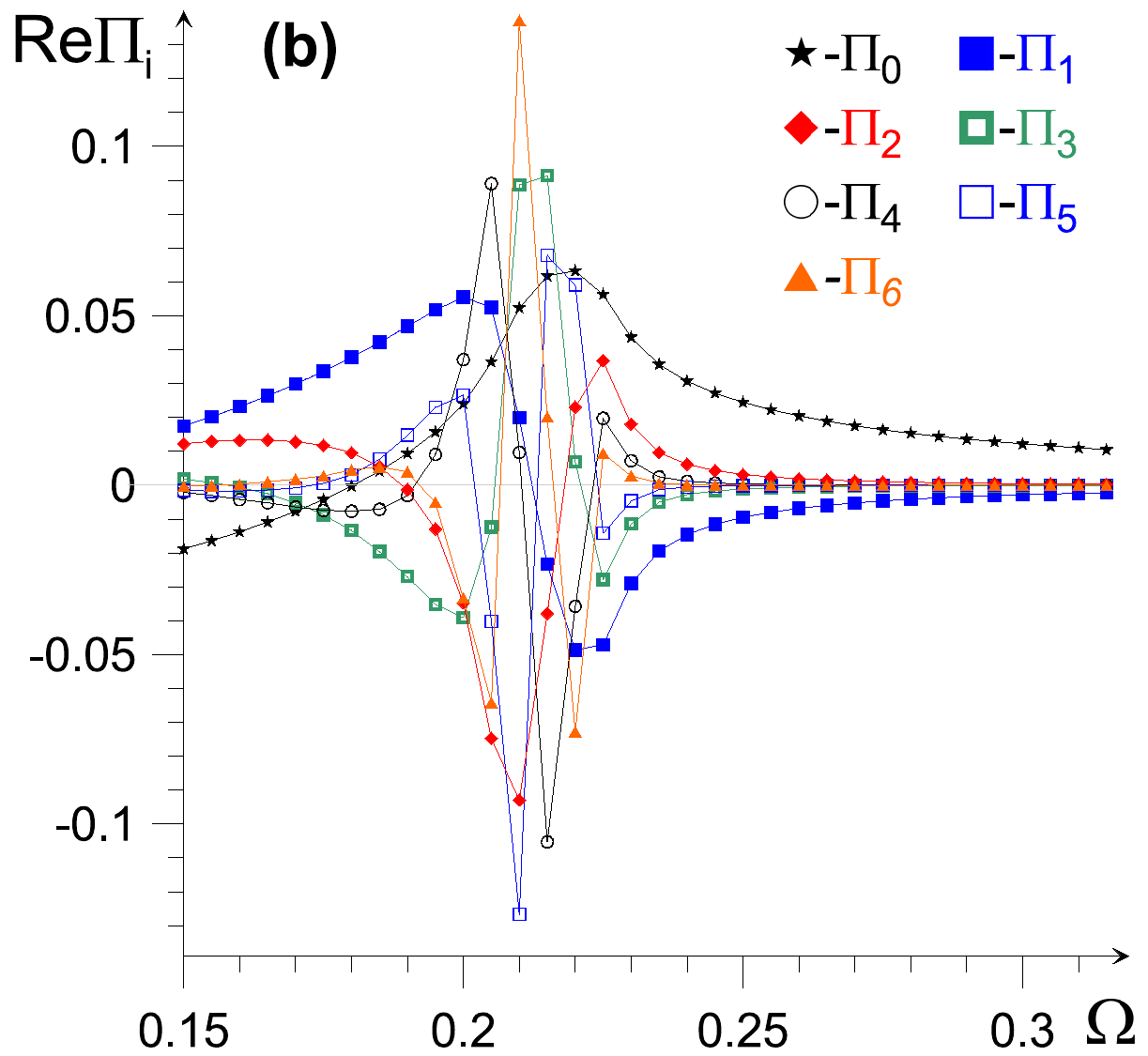}}
\centerline{\includegraphics[angle = 0,width=0.48\columnwidth]{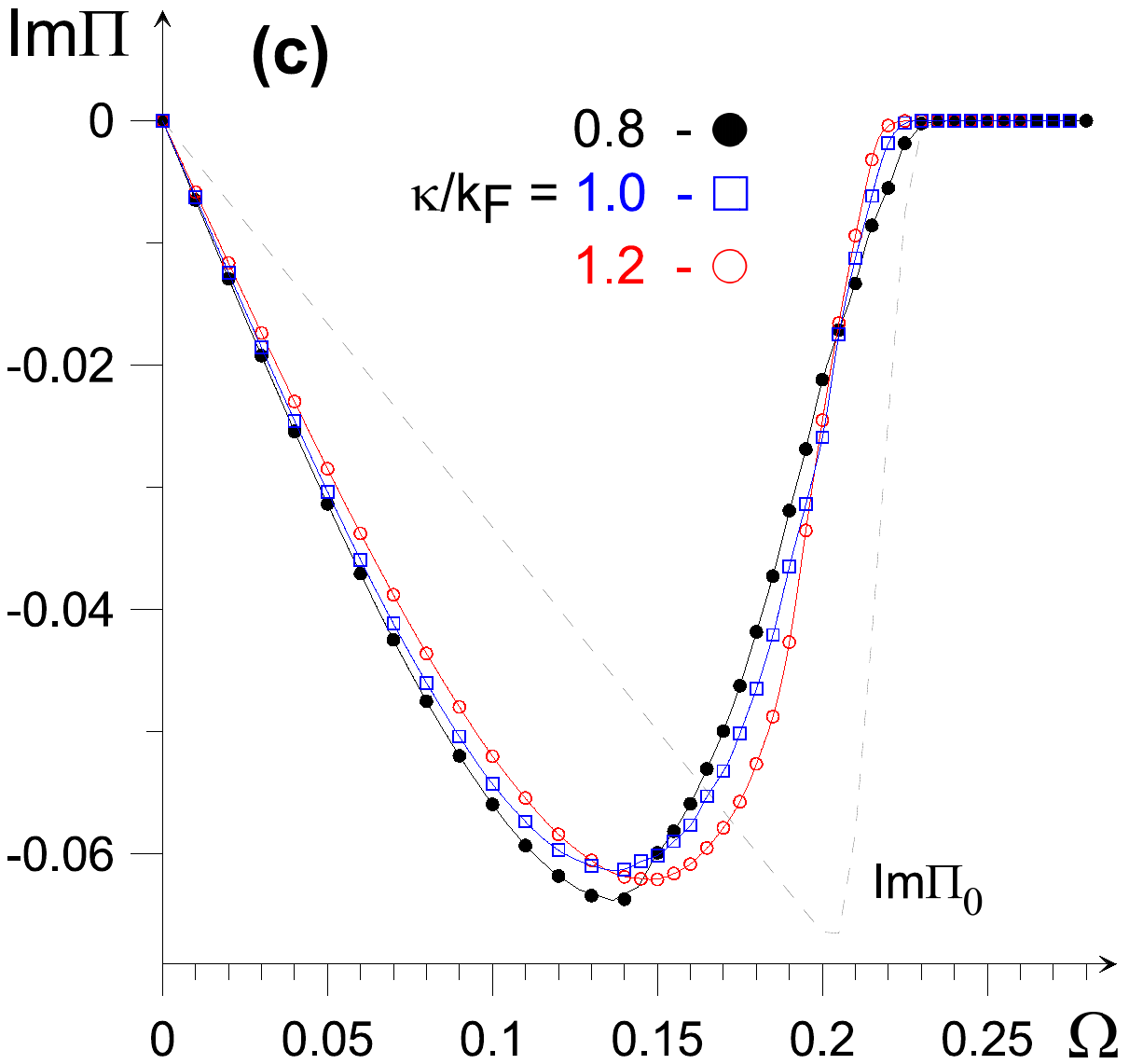}
            \includegraphics[angle = 0,width=0.48\columnwidth]{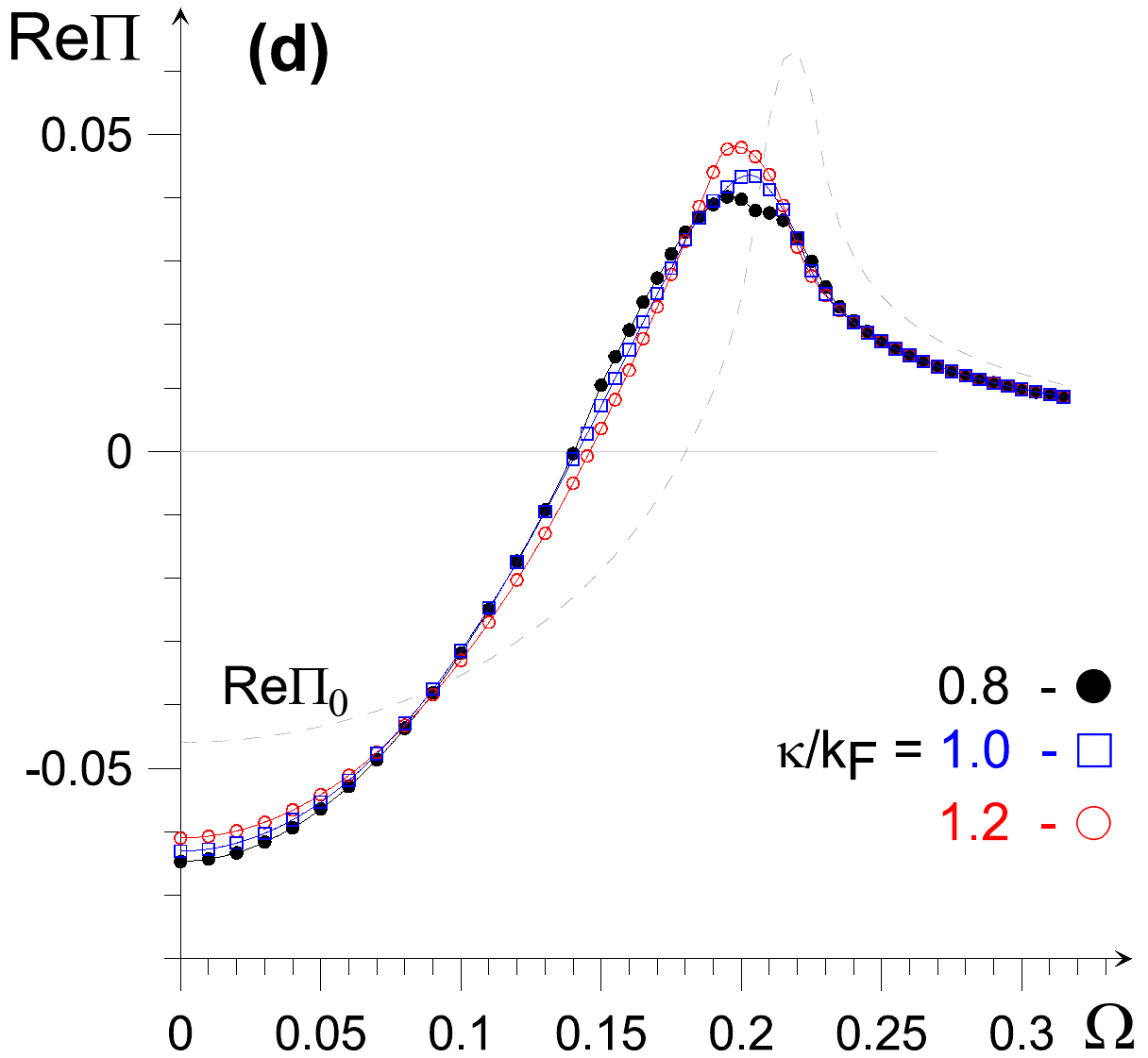}}
\caption{Partial $\textrm{Im} \Pi_i$ (a) and $\textrm{Re} \Pi_i$ (b) contributions to polarization from ladder diagrams with $i$ rungs for $Q/k_F=0.09844$, $r_s=2$, $T/\varepsilon_F=0.02$, and $\kappa/k_F=0.8$. Results for the final answer after series resummation (see text) are shown in panels (c) and (d) for $\textrm{Im} \Pi$ and $\textrm{Re} \Pi$, respectively. The thin dashed lines in panels (c) and (d) are the zeroth-order $\Pi_0$ results for $\kappa/k_F=0.8$. Error bars are smaller than symbol sizes.}
\label{Fig3}
\end{figure}

\begin{figure}[t]
\centerline{\includegraphics[angle = 0,width=0.5\columnwidth]{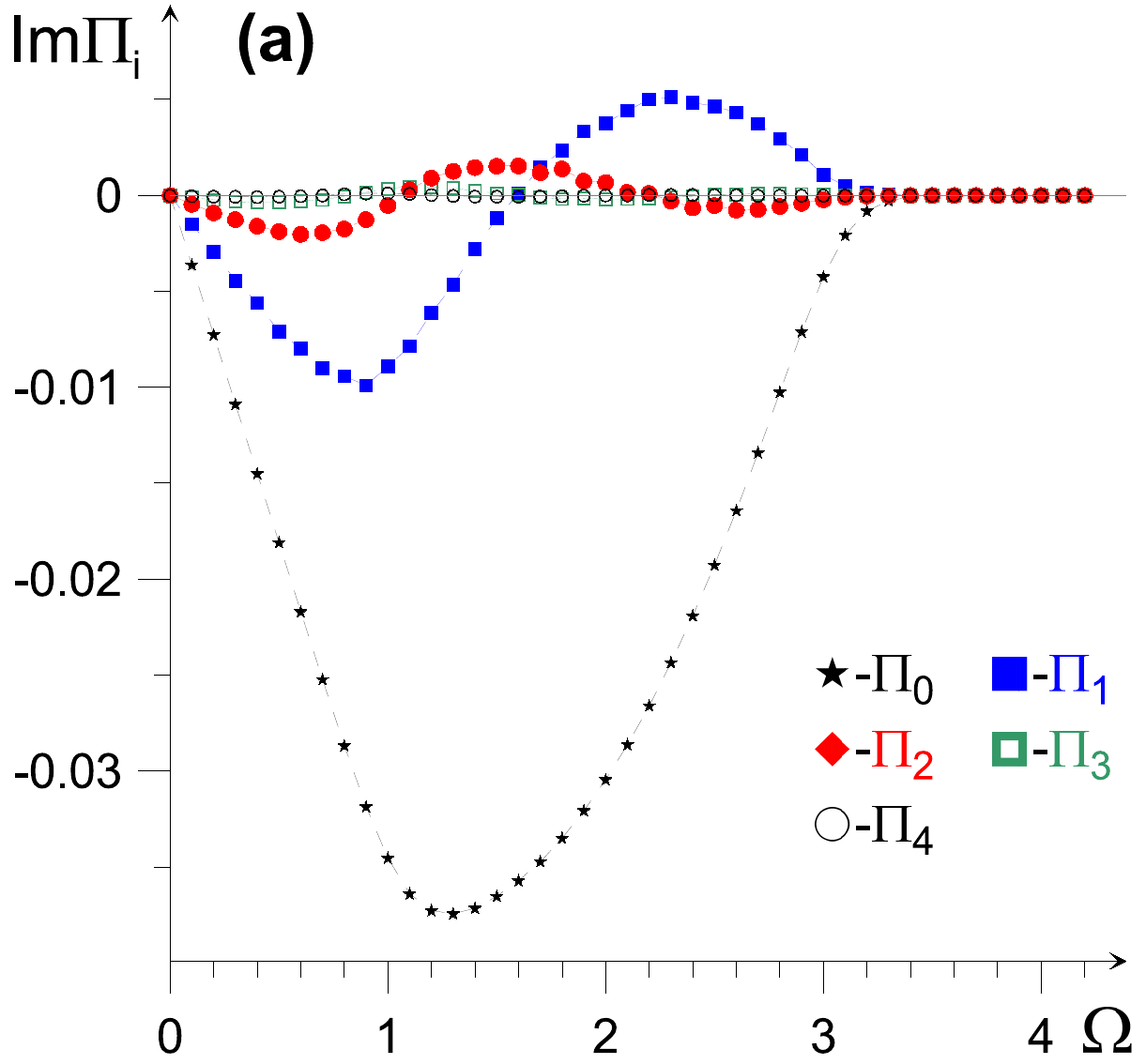}
            \includegraphics[angle = 0,width=0.5\columnwidth]{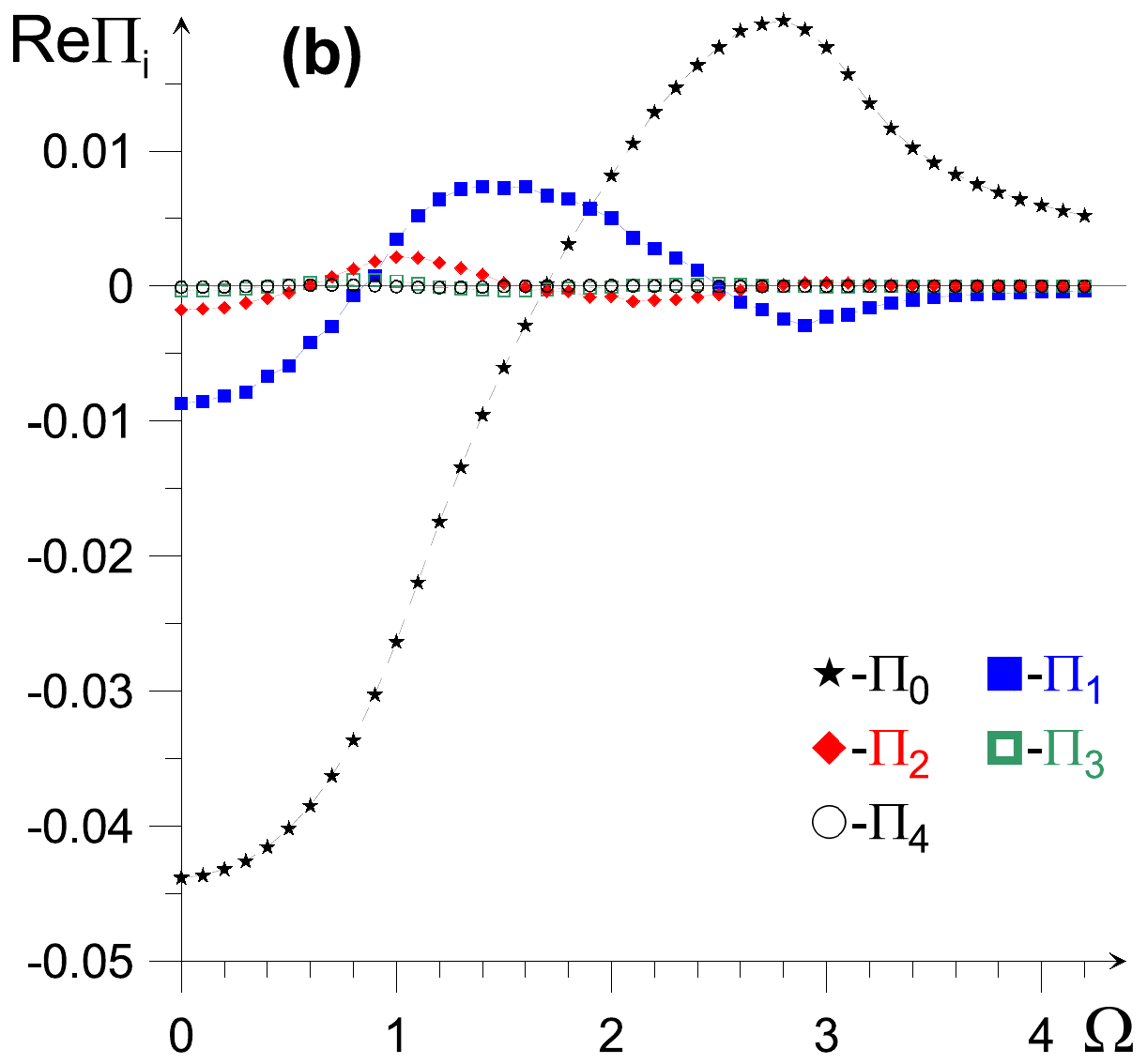}}
\centerline{\includegraphics[angle = 0,width=0.5\columnwidth]{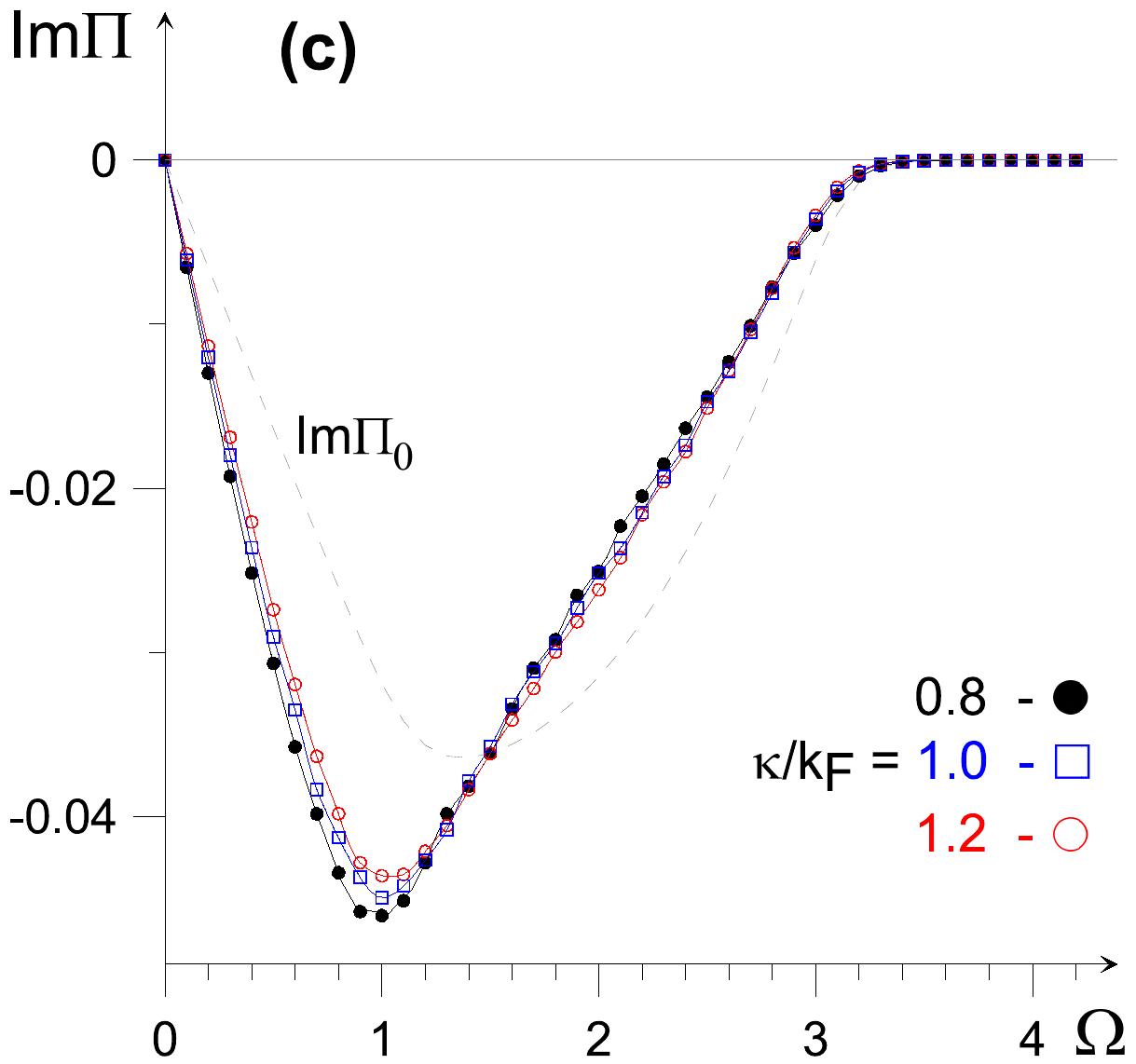}
            \includegraphics[angle = 0,width=0.5\columnwidth]{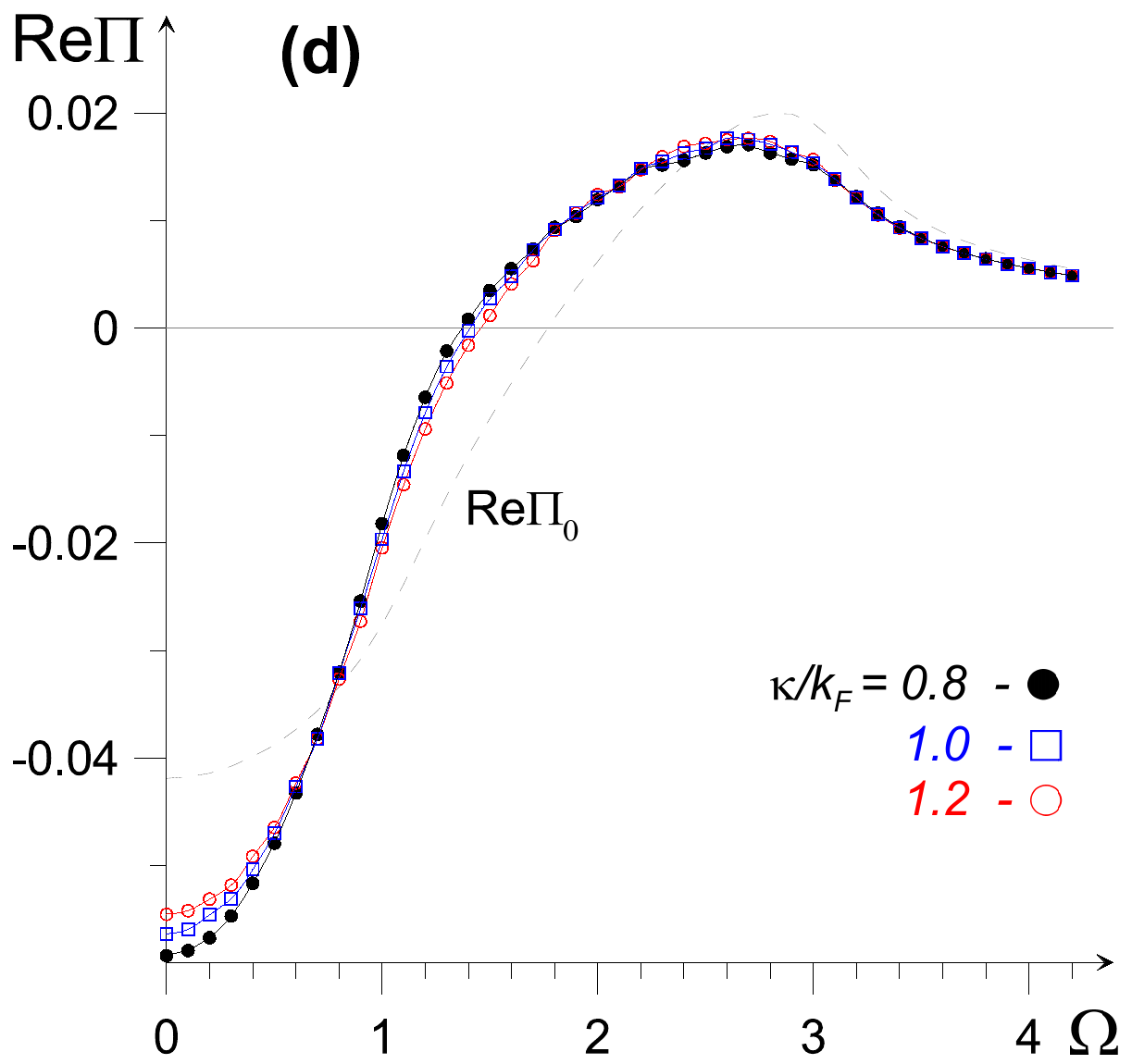}}
\caption{Partial $\textrm{Im} \Pi_i$ (a) and $\textrm{Re} \Pi_i$ (b) contributions to polarization from ladder diagrams with $i$ rungs for $Q/k_F=1.0$, $r_s=2$, $T/\varepsilon_F=0.1$, and $\kappa/k_F=1.2$. Results for the final answer after summing up the series are shown in panels (c) and (d) for $\textrm{Im} \Pi$ and $\textrm{Re} \Pi$, respectively. The thin dashed lines in panels (c) and (d) are the zeroth-order $\Pi_0$ results for $\kappa/k_F=0.8$.  Error bars are smaller than symbol sizes.}
\label{Fig4}
\end{figure}

The convergence of the ladder series strongly depends on $R = \Omega / Qv_F$ ratio with $v_F$ the Fermi velocity, see panels (a) and (b) in Fig.~\ref{Fig3} where we plot partial contributions to polarization $\Pi$ from ladder diagrams in $\Gamma^{(3)}$. For large and small $R$ we observe excellent convergence for all values of $\kappa$ considered in this work. However, in the $R \sim 1$ region the series start diverging for $\kappa \leq 1$, and the problem is getting progressively more severe as $\kappa $ is reduced. At $Q/k_F \approx 0.1$, the series for $\kappa =0.8$ are already visibly beyond the convergence radius. The dimensionless parameter controlling series convergence can be estimated from
\begin{equation}
g \sim \rho_F \frac{4\pi e^2 }{\kappa^2} \sim \frac{\kappa_{TF}^2}{2\kappa^2}
\label{g}
\end{equation}
where $\rho_F$ is the Fermi-energy density of states per spin. When the screening momentum is increased to $\kappa = 1.2$, the series start to converge for any value of $R$ (similarly to what is observed for larger momentum and higher temperature in panels (a) and (b) in Fig.~\ref{Fig4}).

If we formally consider ladder diagrams as Taylor series expansion in powers of parameter $\xi$
with the physical result corresponding to $\xi_0 =1$, see the lower-right block in Fig.~\ref{Fig1}, then series resummation can be performed with the help of the conformal map transformation, $z=\xi f (\xi ) \rightleftarrows \xi= z s (z) \equiv \sum_{m=1}^{\infty} c_m z^m $. By substituting $\xi(z)$ into the expression for $\Pi$ in Fig.~\ref{Fig1}(e), one generates Taylor series in powers of $z$ with the physical result corresponding to $ z_0 = \xi_0 f(\xi_0)$. The map is designed to have $z_0$ within the convergence radius so that the new Taylor series in powers of $z_0$ converge. We find that all our data can be successfully resummed by using a simple-pole map, $z=\xi /(\xi + \xi_p(R))$, with $\xi_p (R=1) \sim 1$ and increasing away from the $R=1$ point. The final results for three values of the Yukawa screening are shown in panels (c) and (d) in Figs.~\ref{Fig3} and \ref{Fig4}. Remarkably, despite substantial differences from the zeroth-order RPA-type results, there is extremely close agreement between the final curves for different values of $\kappa$ (note that $\kappa^2$ was changed by more than a factor of two). Resummation of Taylor series, as opposed to self-consistent and skeleton sequences,
does not lead to multi-valued solutions of the type observed in Ref.~\cite{Kozik15}, not to mention that
resummed and converges "as is" curves end up on top of each other.

At large momentum $Q/k_F=1$ and higher temperature $T/\varepsilon_F$, the series converge well for all values of $\kappa/k_F$ considered and can be summed up ``as is". The final converged results are shown in panels (c) and (d) in Fig.~\ref{Fig4}. The dependence on the Yukawa screening is barely detectable in this parameter regime.
While Figs.~\ref{Fig3} and \ref{Fig4} present results for $r_s=2$, we did not see any notable differences in efficiency of calculations neither for $r_s=1$ nor for $r_s=4$, in agreement with Eq.~(\ref{g}) which predicts that for $\kappa \sim \kappa_{TF}$ one has $g \sim 1$.

Within the RPA, the slope of $\textrm{Im} \Pi \propto \Omega /Q $ dependence is determined by the electron dispersion near the Fermi surface. Multiple scattering changes this result by nearly $100 \%$, which has direct implications for accuracy of Landau damping (and other energy looses in metals) estimates based on the RPA-type calculations (for more details on Landau damping see Ref.~\cite{LD_TP_2023}).

\section{Conclusions}

We presented an efficient technique for computing ladder-diagram series for
polarization on the real frequency axis at finite temperature, or, equivalently, for solving the Bethe-Salpeter equation for the $3$-point vertex function in the ladder approximation and convoluting it with two Green's functions in the Hartree-Fock basis. Our method is numerically exact once all approximations at the level of diagrams involved are specified, i.e. it does not involve the ill-conditioned numerical analytic continuation, finite regularization of the poles, or momentum grids for evaluating multi-dimensional integrals.

Remarkably, final results are nearly independent of the Yukawa screening used to setup the expansion, indicating small bias coming from this auxiliary parameter. We find that multiple scattering of the particle-hole pair results in significant changes of the polarization dependence on the $\Omega /Q$ ratio and, correspondingly, of the Landau damping. We expect that our scheme can be adapted for analogous simulations of realistic materials, and further advanced to include retardation effects for effective interaction potentials.

\section{Acknowledgements} 

This work was supported by the U.S. Department of Energy, Office of Science, Basic Energy Sciences, under Award DE-SC0023141.

\end{document}